\documentclass{midl} 

\usepackage{mwe} 
\jmlrproceedings{MIDL}{Medical Imaging with Deep Learning}
\jmlrpages{}
\jmlryear{2019}
\jmlrworkshop{MIDL 2019 -- Extended Abstract Track}

\title[Short Title]{Effect of Adding Probabilistic Zonal Prior in Deep Learning-based Prostate Cancer Detection }

\midlauthor{\Name{Matin Hosseinzadeh} \Email{Matin.Hosseinzadeh@radboudumc.nl} \\
\Name{Patrick Brand} \Email{Patrick.Brand@radboudumc.nl} \\
\Name{Henkjan Huisman} \Email{Henkjan.Huisman@radboudumc.nl} \\
\addr Diagnostic Image Analysis Group, Radboudumc, Nijmegen, The Netherlands}

\begin{document}

\maketitle

\begin{abstract}
We propose and evaluate a novel method for automatically detecting clinically significant prostate cancer (csPCa) in bi-parametric magnetic resonance imaging (bpMRI). Prostate zones play an important role in the assessment of prostate cancer on MRI. We hypothesize that the inclusion of zonal information can improve the performance of a deep learning based csPCa lesion detection model. However, segmentation of prostate zones is challenging and therefore deterministic models are inaccurate. Hence, we investigated probabilistic zonal segmentation. Our baseline detection model is a 2DUNet trained to produce a csPCa heatmap followed by a 3D detector. We experimented with the integration of zonal prior information by fusing the output of an anisotropic 3DUNet trained to produce either a deterministic or probabilistic map for each prostate zone. We also investigate the effect of early or late fusion on csPCa detection. All methods were trained and tested on 848 bpMRI. The results show that fusing zonal prior knowledge improves the baseline detection model with a preference for probabilistic over deterministic zonal segmentation.

\end{abstract}


\section{Introduction}

Prostate cancer is the second most frequent cancer and the fifth leading cause of cancer death in men \cite{bray2018global}. Early detection of prostate cancer can decrease the mortality rate and make the disease treatable. 
In recent years, prostate MRI has demonstrated the ability of prostate cancer diagnosis and now it is one of the main imaging tools for detecting prostate cancer in clinic\cite{van2019head}. However, diagnosing and grading of prostate cancer lesions using MR images is difficult and requires substantial expertise\cite{rosenkrantz2016interobserver}.

Computer-Aided Detection (CAD) systems can help radiologists by automatically detecting the clinically significant prostate cancer (csPCa) lesions in MR images, but current CAD systems are still performing below the expert level. Prostate CAD systems may be improved by including zonal segmentations, since the prostate zones play a crucial role in diagnostic process in clinic because the occurrence and appearance of prostate cancer are dependent on its zonal location. 
The two main areas of interest in the prostate are the transition zone (TZ) and the peripheral zone (PZ). The PZ is the area where most clinically significant prostate cancer lesions grow, approximately 70\%-75\% of prostate cancers originate in the PZ and 20\%-30\% in the TZ \cite{weinreb2016pi}. Moreover, cancers of these two zones exhibit different behaviors. Based on the location of the lesion, different MRI modalities are majorly used for determining the type of the lesion. For the PZ, DWI is the primary determining sequence, but for the TZ, T2W is primary determining sequence \cite{weinreb2016pi}. However, automatic prostate zonal segmentation is challenging since the boundaries especially at the base, apex and the interface of the zones are usually ambiguous and consequently, the accuracy of automatic segmentations methods especially at PZ is very low (Dice similarity coefficient 0.67) \cite{de2018autoencoders}.

We hypothesize, by incorporating computer-generated prostate zonal probabilistic segmentation as prior knowledge to a deep learning model we can improve csPCa lesion detection performance.

\section{Methods}
\subsection{Data}
Data used in this study was a local, retrospective dataset of consecutive bpMRI (T2W and computed ADC and high b-value DWI) of 848 patients. In this data, lesions were clinically reported by expert radiologists using PIRADSv2 \cite{weinreb2016pi}. As a model of csPCa, all PIRADS 4 and 5 lesions were selected and manually delineated by 2 students. In total 319 patients have at least one csPCa lesion. 
All images were re-sampled to 0.5x0.5x3.6mm resolution and cropped by 9.6x9.6cm around the center. Training, validation and test sets were generated by randomly distributing the patients in a ratio of 3:1:1 through stratified sampling. Thus, they had non-overlapping patients and equal distribution of patients with csPCa was in each set.

\subsection{Experiments}
All networks were trained for 200 epochs using Adam optimizer and weighted cross entropy loss, with a learning rate of $ 10^{-5} $. Data augmentation was applied during the training phase to reduce overfitting.
All predicted 2D heatmaps of a patient were combined to create a 3D volumetric heatmap, on which a two-threshold method was applied to segment csPCa lesions with a probability score for each of them.
For model selection, we selected the best validation model based on the average sensitivity at several points on Free-response ROC (FROC) curves. 
\figureref{fig:overview} gives an overview of the proposed method. 

\textbf{Experiment 1 - Baseline:}
We trained a 2DUnet \cite{ronneberger2015u} with 3 input channels to segment/detect the lesions.

\textbf{Experiment 2 - Early Fusion:}
As the manual prostate zonal segmentations were not available for our dataset we used a modified 3DUnet for anisotropic images \cite{mooij2018automatic}, which were trained using 53 T2W images and manual zonal segmentations, to generate prostate probabilistic and deterministic zonal segmentations for all cases in this study. We used these zonal segmentations as extra channels at the input of the baseline model.

\textbf{Experiment 3 - Late Fusion:}
We did an experiment same as Experiment 2 but instead of using zonal segmentations as inputs of the model, we combined them to the last feature map of the UNet before the 1x1 convolution layer.


\begin{figure}[htbp]
\centering
\floatconts
  {fig:overview}
  {\caption{Schematic overview of the method}}
  {\includegraphics[width=0.9\textwidth]{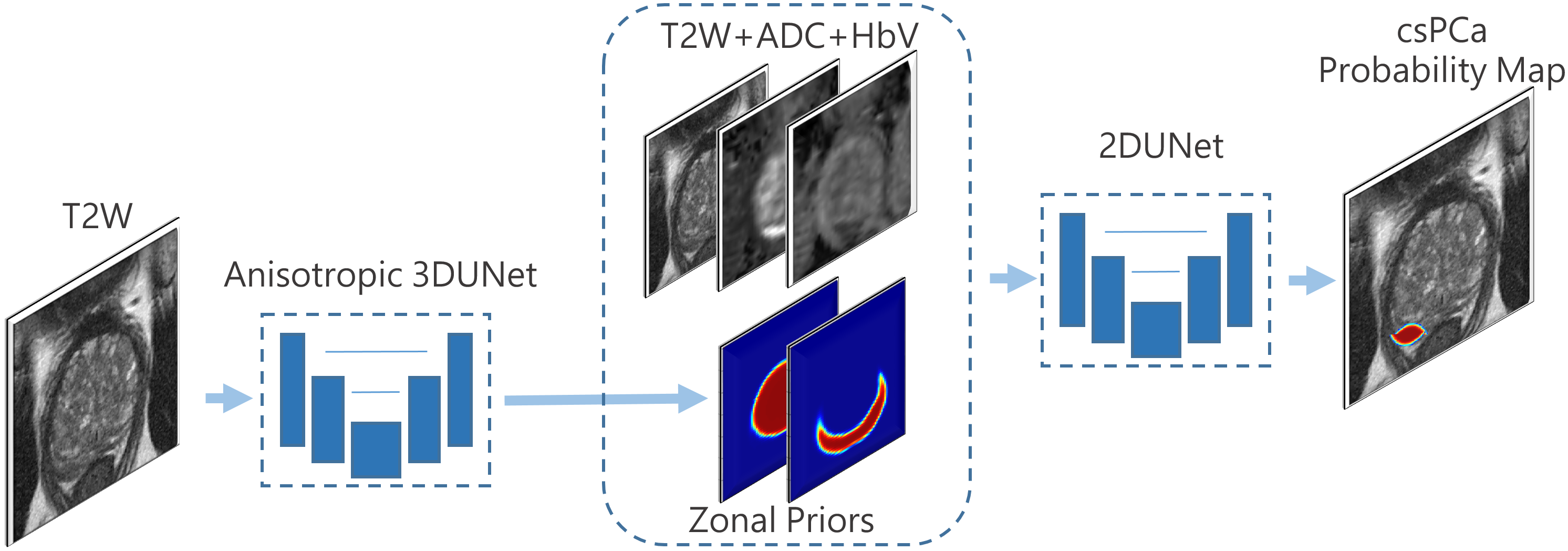}}
\end{figure}

\begin{table}[htbp]
\floatconts
  {tab:Results}%
  {\caption{Sensitivity at various FPs per patient on the test set}}%
  {\begin{tabular}{lllll}
  \bfseries Model & \bfseries Sens@0.5FP & \bfseries Sens@1FP & \bfseries Sens@2FPs & \bfseries Average\\
  Baseline & 0.760 & 0.825 & 0.825 & 0.803\\
  Early Fusion - Probabilistic & \textbf{0.795} & \textbf{0.887} & \textbf{0.887} & \textbf{0.856}\\
  Late Fusion - Probabilistic & 0.774 & 0.873 & 0.873 & 0.840\\
  Early Fusion - Deterministic & 0.774 & 0.802 & 0.802 & 0.793\\
  Late Fusion - Deterministic & 0.774 & 0.816 & 0.816 & 0.802\\
  \end{tabular}}
\end{table}

\section{Results and Discussions}
\tableref{tab:Results} shows that the models trained using probabilistic zonal segmentations achieved better performance on the lesion detection task. Particularly these models find more csPCa lesions in the same FP rate compared to the baseline and deterministic models.

Well trained deep learning networks with sufficient data are meant to automatically learn most useful features in the data such as prostate zones information. However, in the medical imaging domain the data and annotations are scarce and providing abundant data is usually impossible. As a result, providing the deep learning model with prior knowledge can be beneficial. In this paper, we showed that, given the same size of training data, providing prior knowledge using a two-stage approach can help a detection model to perform better compared to a single model which could not capture all prostate zonal information automatically. 
Moreover, we showed that when deterministic segmentation is challenging, probabilistic segmentations can be more beneficial for providing prior knowledge. 


In conclusion the results demonstrate that the proposed early fusion of probabilistic segmentations method achieves the best results among the compared methods in this paper.


\bibliography{hosseinzadeh19}

\begin{thebibliography}{7}
\providecommand{\natexlab}[1]{#1}
\providecommand{\url}[1]{\texttt{#1}}
\expandafter\ifx\csname urlstyle\endcsname\relax
  \providecommand{\doi}[1]{doi: #1}\else
  \providecommand{\doi}{doi: \begingroup \urlstyle{rm}\Url}\fi

\bibitem[Bray et~al.(2018)Bray, Ferlay, Soerjomataram, Siegel, Torre, and
  Jemal]{bray2018global}
Freddie Bray, Jacques Ferlay, Isabelle Soerjomataram, Rebecca~L Siegel,
  Lindsey~A Torre, and Ahmedin Jemal.
\newblock Global cancer statistics 2018: Globocan estimates of incidence and
  mortality worldwide for 36 cancers in 185 countries.
\newblock \emph{CA: a cancer journal for clinicians}, 68\penalty0 (6):\penalty0
  394--424, 2018.

\bibitem[de~Gelder and Huisman(2018)]{de2018autoencoders}
Ard de~Gelder and Henkjan Huisman.
\newblock Autoencoders for multi-label prostate mr segmentation.
\newblock \emph{arXiv preprint arXiv:1806.08216}, 2018.

\bibitem[Mooij et~al.(2018)Mooij, Bagulho, and Huisman]{mooij2018automatic}
Germonda Mooij, Ines Bagulho, and Henkjan Huisman.
\newblock Automatic segmentation of prostate zones.
\newblock \emph{arXiv preprint arXiv:1806.07146}, 2018.

\bibitem[Ronneberger et~al.(2015)Ronneberger, Fischer, and
  Brox]{ronneberger2015u}
Olaf Ronneberger, Philipp Fischer, and Thomas Brox.
\newblock U-net: Convolutional networks for biomedical image segmentation.
\newblock In \emph{International Conference on Medical image computing and
  computer-assisted intervention}, pages 234--241. Springer, 2015.

\bibitem[Rosenkrantz et~al.(2016)Rosenkrantz, Ginocchio, Cornfeld, Froemming,
  Gupta, Turkbey, Westphalen, Babb, and Margolis]{rosenkrantz2016interobserver}
Andrew~B Rosenkrantz, Luke~A Ginocchio, Daniel Cornfeld, Adam~T Froemming,
  Rajan~T Gupta, Baris Turkbey, Antonio~C Westphalen, James~S Babb, and
  Daniel~J Margolis.
\newblock Interobserver reproducibility of the pi-rads version 2 lexicon: a
  multicenter study of six experienced prostate radiologists.
\newblock \emph{Radiology}, 280\penalty0 (3):\penalty0 793--804, 2016.

\bibitem[van~der Leest et~al.(2019)van~der Leest, Cornel, Israel, Hendriks,
  Padhani, Hoogenboom, Zamecnik, Bakker, Setiasti, Veltman,
  et~al.]{van2019head}
Marloes van~der Leest, Erik Cornel, Bas Israel, Rianne Hendriks, Anwar~R
  Padhani, Martijn Hoogenboom, Patrik Zamecnik, Dirk Bakker, Anglita~Yanti
  Setiasti, Jeroen Veltman, et~al.
\newblock Head-to-head comparison of transrectal ultrasound-guided prostate
  biopsy versus multiparametric prostate resonance imaging with subsequent
  magnetic resonance-guided biopsy in biopsy-na{\"\i}ve men with elevated
  prostate-specific antigen: a large prospective multicenter clinical study.
\newblock \emph{European urology}, 75\penalty0 (4):\penalty0 570--578, 2019.

\bibitem[Weinreb et~al.(2016)Weinreb, Barentsz, Choyke, Cornud, Haider, Macura,
  Margolis, Schnall, Shtern, Tempany, et~al.]{weinreb2016pi}
Jeffrey~C Weinreb, Jelle~O Barentsz, Peter~L Choyke, Francois Cornud, Masoom~A
  Haider, Katarzyna~J Macura, Daniel Margolis, Mitchell~D Schnall, Faina
  Shtern, Clare~M Tempany, et~al.
\newblock Pi-rads prostate imaging--reporting and data system: 2015, version 2.
\newblock \emph{European urology}, 69\penalty0 (1):\penalty0 16--40, 2016.

\end{thebibliography}

\end{document}